\documentclass[sigconf,nonacm]{acmart}

\AtBeginDocument{}

\setcopyright{acmlicensed}
\copyrightyear{2024}
\acmYear{2024}
\acmDOI{XXXXXXX.XXXXXXX}

\usepackage{subcaption}
\usepackage{spverbatim}

\begin{document}
\title{An Efficient and Explanatory Image and Text Clustering System with Multimodal Autoencoder Architecture}

\begin{comment}
\author{Anon}
\email{.}
\affiliation{
    \institution{.}
    \city{.}
    \state{.}
    \country{.}
    \postcode{.}
}
\end{comment}

% \begin{comment}
\author{Tiancheng Shi}
\authornote{Both authors contributed equally to this research.}
\email{ts3474@columbia.edu}
\affiliation{
    \institution{Columbia University}
    \city{New York}
    \state{NY}
    \country{USA}
    \postcode{10027}
}

\author{Yuanchen Wei}
\authornotemark[1]
\email{yw3939@columbia.edu}
\affiliation{
    \institution{Columbia University}
    \city{New York}
    \state{NY}
    \country{USA}
    \postcode{10027}
}

\author{John R. Kender}
\email{jrk@cs.columbia.edu}
\affiliation{
    \institution{Columbia University}
    \city{New York}
    \state{NY}
    \country{USA}
    \postcode{10027}
}
% \end{comment}


\begin{abstract}
We demonstrate the efficiencies and explanatory abilities of extensions to the common tools of Autoencoders and LLM interpreters, in the novel context of comparing different cultural approaches to the same international news event.
We develop a new Convolutional-Recurrent Variational Autoencoder (CRVAE) model that extends the modalities of previous CVAE models, by using fully-connected latent layers to embed in parallel the CNN encodings of video frames, together with the LSTM encodings of their related text derived from audio.
We incorporate the model within a larger system that includes frame-caption alignment, latent space vector clustering, and a novel LLM-based cluster interpreter.
We measure, tune, and apply this system to the task of summarizing a video into three to five thematic clusters, with each theme described by ten LLM-produced phrases.
We apply this system to two news topics, COVID-19 and the Winter Olympics, and five other topics are in progress.
\end{abstract}

\begin{comment}
\begin{CCSXML}
<ccs2012>
   <concept>
       <concept_id>10010147.10010178.10010179.10003352</concept_id>
       <concept_desc>Computing methodologies~Information extraction</concept_desc>
       <concept_significance>500</concept_significance>
       </concept>
   <concept>
       <concept_id>10010147.10010178.10010224.10010240</concept_id>
       <concept_desc>Computing methodologies~Computer vision representations</concept_desc>
       <concept_significance>500</concept_significance>
       </concept>
   <concept>
       <concept_id>10010147.10010257.10010293.10010294</concept_id>
       <concept_desc>Computing methodologies~Neural networks</concept_desc>
       <concept_significance>500</concept_significance>
       </concept>
   <concept>
       <concept_id>10010405.10010469.10010474</concept_id>
       <concept_desc>Applied computing~Media arts</concept_desc>
       <concept_significance>500</concept_significance>
       </concept>
 </ccs2012>
\end{CCSXML}

\ccsdesc[500]{Computing methodologies~Information extraction}
\ccsdesc[500]{Computing methodologies~Computer vision representations}
\ccsdesc[500]{Computing methodologies~Neural networks}
\ccsdesc[500]{Applied computing~Media arts}

\keywords{multimodal learning, representation learning, multimedia understanding, cross-culture comparison}
\end{comment}


\maketitle

% ==================================================

\section{Introduction}

Videos convey information with high temporal density through both images and audio.
Thus, video content extraction and abstraction methods are gradually gaining in their significance.
However, existing common tools are costly to train and opaque to interpret.
We explore two novel variations on Autoencoders and LLM interpreters to generate thematic clusters of video tags without human supervision, in order to summarize individual videos.
Our final goal is to quickly compare the videos of the same event presented by different cultures, in order to gain insight into their differing perspectives.

\begin{figure*}
    \centering
    \begin{subfigure}{0.48\textwidth}
        \centering
        \includegraphics[scale=0.24]{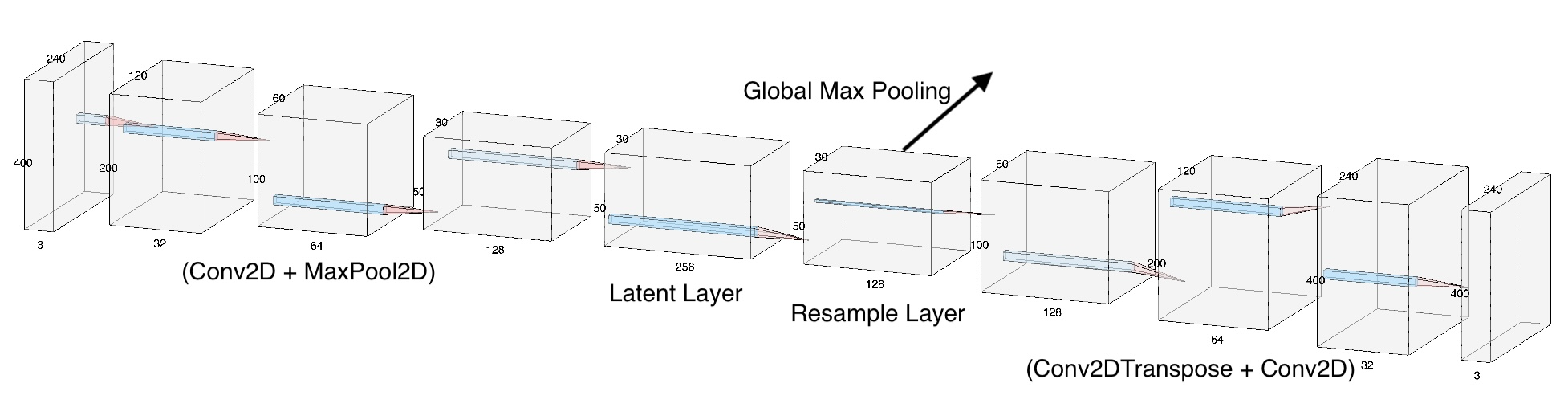}
        \caption{Pure CVAE consists only of convolution layers, as represented by “boxes”, of which the widths shows the number of channels in a layer.}
        \label{subfig:pure_cvae}
    \end{subfigure}%
    \hfill
    \begin{subfigure}{0.48\textwidth}
        \centering
        \includegraphics[scale=0.24]{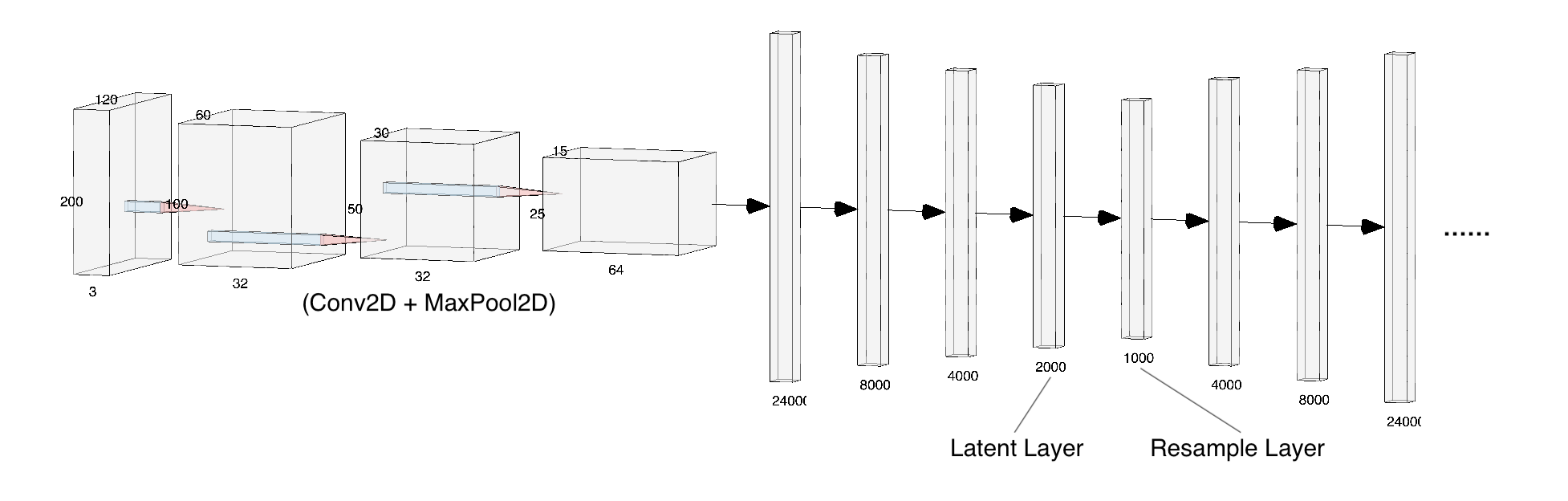}
        \caption{Dense CVAE consists of convolution layers in the front and end, and fully-connected layers in the middle, as represented by “columns” (duplicated convolution layers at the end are omitted).}
        \label{subfig:dense_cvae}
    \end{subfigure}
    \caption{Pure and dense CVAE Architectures.}
    \label{fig:pure_and_dense}
    \Description{Two CVAE architectures. The left one has several convolution layers, and the right one has convolution and fully connected layers.}
\end{figure*}

Video frames with high resolution are associated with an extremely high dimension that makes clustering difficult.
Prior work has inspired us that Variational Autoencoders with only convolution layers (Pure CVAE, see Figure \ref{subfig:pure_cvae}) can reduce the dimensionality of images low enough for other Machine Learning algorithms to handle them.
However, we note that the use of the Global Max Pooling layer on top of the latent layer undermines the effectiveness of the original CVAE model.
We substitute instead multiple fully-connected linear layers, also following the encoder-decoder structure (Dense CVAE, see Figure \ref{subfig:dense_cvae}), which learn important features independent of the format of input vectors.

We claim the following advantages of our methodology.

\begin{figure*}
  \includegraphics[width=\textwidth]{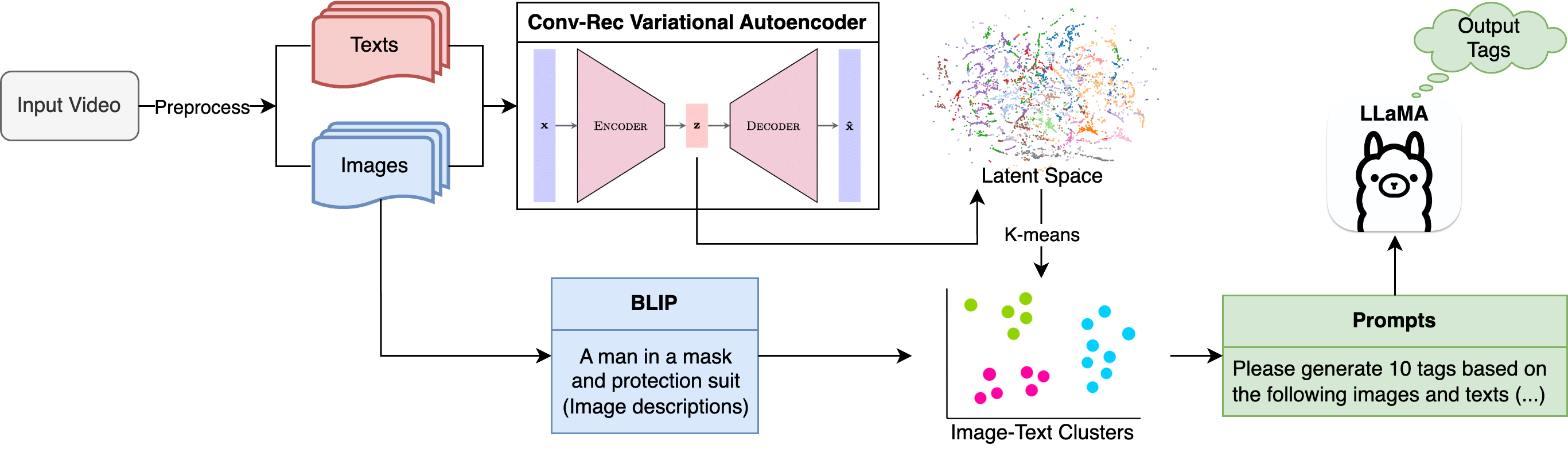}
  \caption{System Overview: CRVAE model encodes images and texts into a low-dimension latent space, the vector representations are clustered using $K$-means, and BLIP describes the video frames to help LLaMA summarize the frames and captions.}
  \label{fig:pipeline}
  \Description{System overview's pipeline details}
\end{figure*}

\paragraph{\textbf{Novelty}}
We propose the CRVAE model that further introduces multimodality by incorporating NLP elements.
While traditional CVAE solely processes images, CRVAE can encode audio information in the original videos, allowing to learn denser information representation comparing with its unimodal counterparts.
Additionally, the mostly automatic full pipeline (see Figure \ref{fig:pipeline}) includes a Large Language Model (LLM) interpretation.
Images from each thematically related cluster, each augmented with a caption generated by the BLIP model, are fed into the LLaMA model, with a prompt asking it to generate a fixed number of tags for the cluster.

\paragraph{\textbf{Universality}}
Our method is applicable to YouTube news videos, regardless of their topics, sources, languages, cultural backgrounds, etc.
This universality is because the key parts of our system---both the CRVAE model and the LLM---are designed to inherently handle multilingual text inputs.

\paragraph{\textbf{Efficiency}}
First, our system is almost fully automatic:
the only manual parameter into the pipeline is the selection of optimal number of clusters for $K$-means, heuristically chosen with the help of metrics and plots.
Secondly, the training session is also cost-efficient enough to be deployed on local PCs.
The CRVAE model does not require prior information, and the training then takes less than 30 minutes per video.
The LLM interpretation phase costs similarly.

\begin{comment}
We claim the following contributions.

\textbf{Architecture.}
We note that the use of the Global Max Pooling layer on top of the latent layer undermines the effectiveness of the original CVAE model.
We substitute instead multiple fully-connected linear layers, also following the encoder-decoder structure (Dense CVAE, see Figure \ref{subfig:dense_cvae}), which learn important features independent of the format of input vectors. 
This allows us to incorporate NLP elements, to handle both texts and images.
Since the audio channels in news videos mainly consist of clear speech of organized sentences, we use ASR to directly transform the audio data into natural languages.

\textbf{Summarization system.}
We present a mostly automatic full pipeline (see Figure \ref{fig:pipeline}) that first applies $K$-means to latent space vectors, producing thematically related clusters.
The images of each cluster, each augmented with a caption generated by the BLIP model, are fed to the LLaMA model, with a prompt asking it to generate a fixed number of tags for the cluster.

\textbf{Cross-cultural video comparisons.}
The above pipeline is tested on two news videos from both Western (English) and Eastern (Chinese) sources.
We visualize the results, noting the several unanticipated image-to-text associations that result.
\end{comment}

\section{Related Works}

\subsection{Convolutional Variational Autoencoder}

Prior work has used the Convolutional Variational Autoencoder (CVAE) model for representation learning and clustering of images (video frames)~\cite{CVAE}.
Specifically, 1-strided convolution layers, each followed by a $2 \times 2$ max pooling layer can be applied to extract low-dimension features from input images.
Yet for images with high resolution, these layers are insufficient to reduce the high dimensions to an approachable level.
Conversely, using larger strides and/or pool sizes will result in fuzziness during reconstruction.

One method to resolve this dilemma was to apply a Global Max Pooling to each channel, further reducing the dimension from $C \times H \times W$ to just $C \times 1$, where $C, H, W$ represents number of channels, image height, and image width respectively. 
However, this brute-force method fails for two reasons.

On the one hand, by forcing a full, 2D feature map to compress to a single value, we lose all spatial information.
This simply keeps track of an extremely “bright spot”, the pixel that is most activated by the convolution layers.
On the other hand, during the actual training process, the latent vector after Global Max Pooling is not involved.
It is neither the output optimized by the Encoder network, nor can it be used to reconstruct the output image by the Decoder network.

We conclude that pure CVAE is insufficient to encode high-resolution images.
In contrast, multiple external researchers (\citet{pu_cvae, fan_cvae}, and others) have use of dense layers between the convolution layers and the latent layer.
Therefore, we adopt the Dense CVAE model: CVAE with dense (fully-connected) layers in the middle; see Figure \ref{subfig:dense_cvae}.

\subsection{Multimodal Representation Learning}

MRL reduces the dimensionality of heterogeneous data, aiming at retaining most information and reducing the gap between different modalities of the input data.
\citet{guo2019deep} classified popular frameworks of Multimodal Representation Learning into three potentially overlapping types.

The first, Joint Representation, method typically involves a low-dimension subspace containing the fusion of information from all modalities. 
A common approach is to use an additive method: $z = f(w_1 v_1 + w_2 v_2)$, where $v_i$ are unimodal representations, $f$ is the activation function, and $z$ is the output multimodal representation.

The second, Coordinated Representation, in contrast learns each modality in separate models and exerts additional constraints to regularize the training across all modalities.

The third, Encoder-Decoder Framework, translates one modality into the other, and takes the intermediate layer as the multimodal representation.

Our work uses a modality-fusion method within an encoder-decoder framework and can be viewed as a mix of the first and third types.
The advantage of this structure is that a joint representation outputs a single vector representation that contains shared information across modalities, and enables clustering and interpretation in the next steps.

\subsection{Bimodal Deep Autoencoder}

\citet{ngiam2011multimodal} proposed a bimodal Autoencoder in a Joint Representation structure.
The model takes both images and audio from videos, fuses them into a shared representation layer, and reconstructs both modalities.

The autoencoder was trained to “denoise” the video: one of the modalities of the input data would be masked to zero at the input, even while both modalities were still evaluated at the output.
This emphasizes the extraction of cross-modality features.
The work showed that the latent layer of this autoencoder structure was an effective low-dimension representation of information from both modalities.

Our model is designed in a similar structure, but trained toward our
goal of learning the shared representation of videos for further analyses.
Consequently, the zero-masks in their work are not included in our experiments.

\section{Methods}

\begin{figure}[ht]
    \centering
    \includegraphics[width=\linewidth]{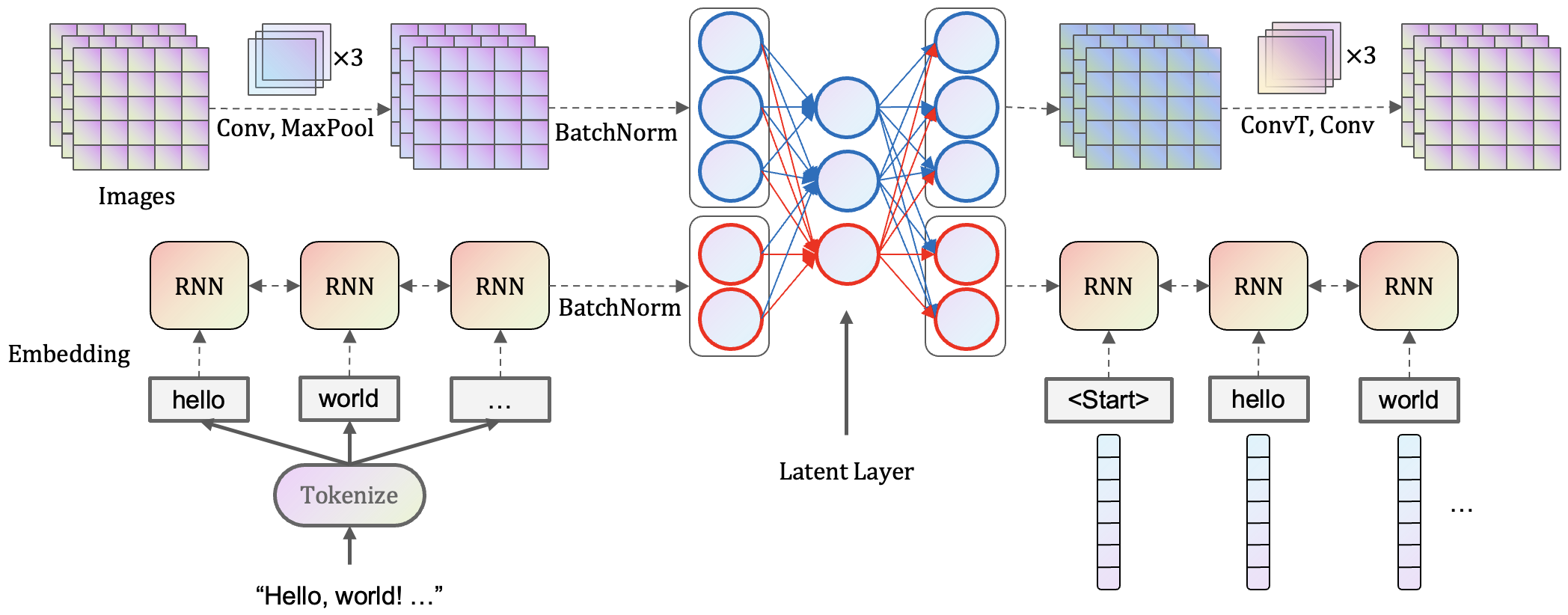}
    \caption{CRVAE Architecture: the encoder, in parallel, uses CNN layers to encode images and uses RNN layers to encode texts. These are fused into fully-connected layers with shrinking size. The decoder has a symmetric structure.}
    \label{fig:crvae}
    \Description{An autoencoder structure with a symmetric encoder and decoder. Each part has several CNN and RNN layers in parallel, and the final layers are fused into several fully connected layers.}
\end{figure}

We propose our new CRVAE (Convolutional-Recurrent Variational Autoencoder) model architecture as shown in Figure \ref{fig:crvae}, which takes an image and a sentence as input each time, processes them in parallel using CNN and LSTM respectively, and combines the multimodal input vectors through fully-connected layers.
It generally follows the encoder-decoder structure, determined by a latent layer in the center.
The final takeaway of this model is the vectors in the latent space, with lower dimensions compared to the input images and text captions.

\subsection{Encoder}

\subsubsection{Images.}
Similar to the dense CVAE model, we set the dimension of input images as $C \times H \times W = 200 \times 120 \times 3$, where 3 represents the RGB channels, and set a batch size of 16.
The framework of Convolution and Max Pooling layers remains the same, only that the numbers of filters (channels) are set to be uniformly 32.
The motivation to use fewer channels in the last convolution layer (decreased from 64 to 32) is that after flattening, we will have $32 \times 25 \times 15 = 12,000$ neurons, instead of 24,000, which significantly reduces the model size.

\subsubsection{Text.}

The text encoder network is an RNN-based model, with the input of a sequence of embedded text in dimension 200.
However, the pre-trained embedding weights (Chinese or English) are fixed, and are not optimized during gradient descent.
The model is selected from either vanilla RNN or LSTM, both of which have a hidden state vectors of length 512 and 2 stacked layers of bidirectional recurrent cells.
Our experiments show that the LSTM model has higher performance than vanilla RNN on all tasks.

\subsubsection{Latent Layers.}
CRVAE uses 2 fully-connected neural layers in the middle.
The outputs of the Convolution layers and LSTM layers are flattened and normalized using the Batch Normalization layer, to ensure that they are approximately on the same scale.
The resulting vector, after concatenation, has the dimension of $32 \times 25 \times 15 + 2 \times 2 \times 512 = 14,048$, where the $2 \times 2$ represents 2 layers of LSTM, and both of them in 2 directions. 
The vector is then narrowed down to 1 layer with 4,000 neurons.
The final latent layer has 2,000 neurons, 1,000 for the latent mean and 1,000 for the latent standard deviation (in log scale).

\subsubsection{Resampling.}
The latent mean $\mu$ and latent standard deviation $\ln \sigma$ are then used to resample (using a Normal distribution) across the latent space.
The resulting vector is passed to the decoder network for the reconstruction of images and text.
After the training session, the latent means $\mu$ are finally produced as CRVAE output.

\subsection{Decoder}
\label{subsec:decoder}

The decoder network is roughly symmetric to the encoder network, as the 1000-dimension embedding is gradually rebuilt to 4000- and then 14,048-dimension.
It is then spit into two parts, and trained to reconstruct the images and text separately.

\subsubsection{Images.}

The image decoder uses Transposed 2D Convolution layers with a $3 \times 3$ kernel and a stride of 2 to “upsample” an image channel, resulting in doubled width and height.
Each of these Transposed Convolution layers is followed by a Convolution layer with the same kernel size ($3 \times 3$) and number of channels (32).
After three of these paired Transposed Convolution and Convolution blocks, the tensor is reconstructed into the original resolution with 3 channels, representing the Red, Green, and Blue pixels of an image.
We use pixel-wise Mean Squared Error (MSE) between the input image and the reconstructed images as the loss function.

\subsubsection{Text.}

The text decoder in CRVAE is different from the traditional LSTM decoder for NLP tasks because we require the model to handle multilingual inputs.
Normally, we would have mapped the neurons to a layer with the same length as the vocabulary size, and assigned a SoftMax activation corresponding to the Cross-Entropy loss.
However, due to the great differences between Chinese and English, this approach does not work well.
Instead, we now require the decoder to predict the text embeddings as tensors, and optimize the MSE loss between the original text and the reconstructed embedded text.
To make this work, we “freeze” the text embedding layer in the encoder; text embeddings are not involved in the training session. 
This necessitates two further changes to the decoder architecture.  

\subsubsection{Teacher Forcing.}
    
We applied the Teacher Forcing algorithm to our text decoder network.
At each LSTM cell, we predict the next word by the input of the \textit{ground-truth} previous word embeddings, instead of the \textit{predicted} previous embedding.
We experienced a significant performance increase with this algorithm, mostly because our purpose is simply to learn important features and characteristics of the text, information which is encoded in the latent layer, and passed to the decoder through hidden and cell states.

\subsubsection{Nearest Neighbors.}

Because our final reconstructions are tensors instead of words, we still need a method to “verbalize” them.
We therefore search for the nearest neighbor of an embedded word in the vocabulary, and “decode” this vector as that word.
Note that this nearest neighbor verbalization is not part of the training process.

\subsection{CRVAE Model Configuration}
\label{subsec:losses}

By constructing our model based on the architecture in Figure~\ref{fig:crvae}, we can formulate two  types of MSE loss: image loss and text loss.
The final loss function is $L = ImageLoss + \lambda \cdot TextLoss$, where $\lambda$ is a ratio hyperparameter to balance the reduction of losses during the training session. In practice, we set $\lambda = 3$.

All intermediate layers in the model are activated by ReLU (Rectified Linear Unit) function.
The AdamW (Adaptive Momentum with Decoupled Weight Decay Regularization) optimizer with a learning rate of $\alpha = 10^{-4}$ is used, and for each parameter, the gradient is clipped to 0.01 to prevent gradient explosion.
The model is trained for 500 epochs on a local device, an NVIDIA GeForce RTX 3080 with 10 GB GPU Memory.
A typical training session lasts for around 30 minutes, which is similar to the training time of a 300-epoch dense CVAE model.

\subsection{Clustering and Cluster Interpretation}

After encoding each <frame, caption> pair into a latent space of dimension 1000, we perform $K$-means clustering on all vectors generated from a video.
To find the best $K$, which typically will be a small integer, we semi-automatically evaluate each trail $K$ using a number of cluster quality metrics.

\subsubsection{Metrics.}

The metrics used include average inter-cluster distance, average cross-cluster distance, and a cluster robustness test.
Specifically, we seek an ideal $K$ that has small inter-cluster distances (which indicate compact clusters), large cross-cluster distances (which indicate distinct clusters), and clusters that do not change significantly as new centroids are introduced.
We plot and visualize these metrics against different $K$, in order to heuristically choose the optimal number of clusters.

\subsubsection{Tags.}
Besides these quantifiable metrics, we also propose a more intuitive
evaluation of cluster quality, in two steps, with each step using an LLM. 

We first ask the BLIP-base model on HuggingFace, \textsc{blip-image-captioning-base} in full \textsc{float32} precision, to “describe” each input image frame using its own captioning service.
This takes only a short prompt (see Appendix \ref{sec:blip_prompt}).

We then use the simplest Llama-2 model on HuggingFace, \textsc{Llama-2-7b-chat-hf}, to perform generative language modeling.
Because text elements for a given thematic cluster can be easily concatenated, both in syntax and semantics, we combine all of a cluster's actual video captions into one textual grouping, and all of its BLIP descriptions into another.
The required prompt is considerably longer and more sensitive to tuning (see Appendix \ref{sec:llama_prompt}).

\section{Dataset}
\label{sec:dataset}

To evaluate our pipeline, we focused on seven news events that arouse global interest, including COVID-19, Winter Olympics 2022, China's “visa-free” visitor policy, Copa America 2024 Soccer Tournament, Beryl Hurricane, China's newly deployed self-driving taxis, and the crisis of cooking oil being contaminated by petroleum in oil trucks in China.
For each topic, we collected one video from English sources and another video from Chinese sources.
We have run experiments on COVID-19 and Winter Olympics, and
are processing the others.

\begin{comment}
To evaluate our pipeline, we focused on two pairs of news videos related to COVID-19 and the Winter Olympics 2022.
We were interested in detecting potential cultural differences which would be reflected in news agency reports.
\end{comment}
Raw image datasets about COVID-19 are sampled at the rate of 1 frame every 2 seconds, while those about Winter Olympics at the rate of 1 frame every 2.5 seconds, in order to balance the data sizes.
We align segments of text data (taken from the audio) with each image frame automatically (for English) or semi-automatically (for Chinese).

% new figure is 1 x 4
\begin{figure}[ht]
    \centering
    \begin{subfigure}{0.1\textwidth}
        \centering
        \includegraphics[scale=0.025]{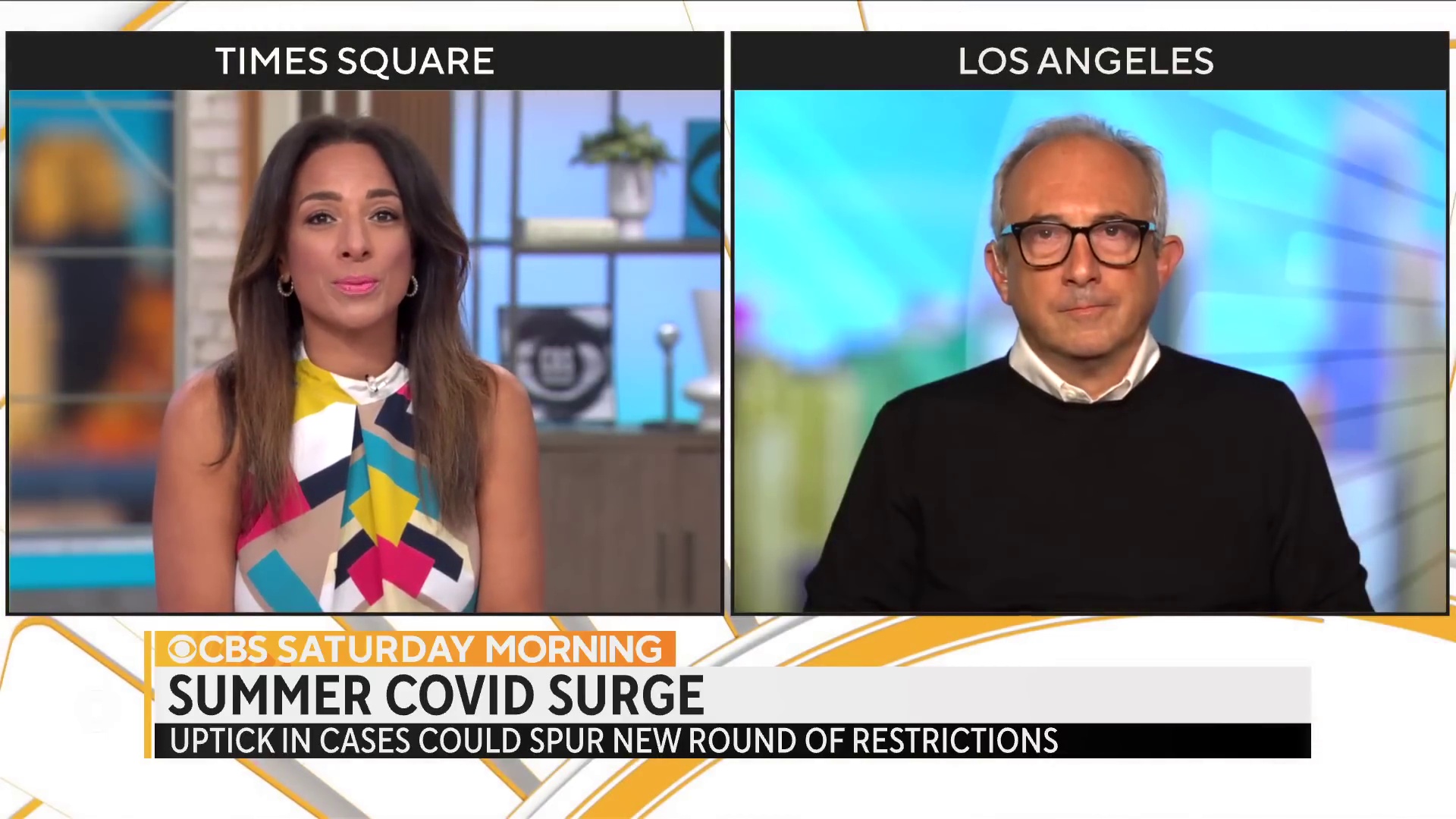}%
        \caption{COVID-19 (English).}
        \label{subfig:en_covid_15}%
    \end{subfigure}
    \hfill
    \begin{subfigure}{0.1\textwidth}
        \centering
        \includegraphics[scale=0.025]{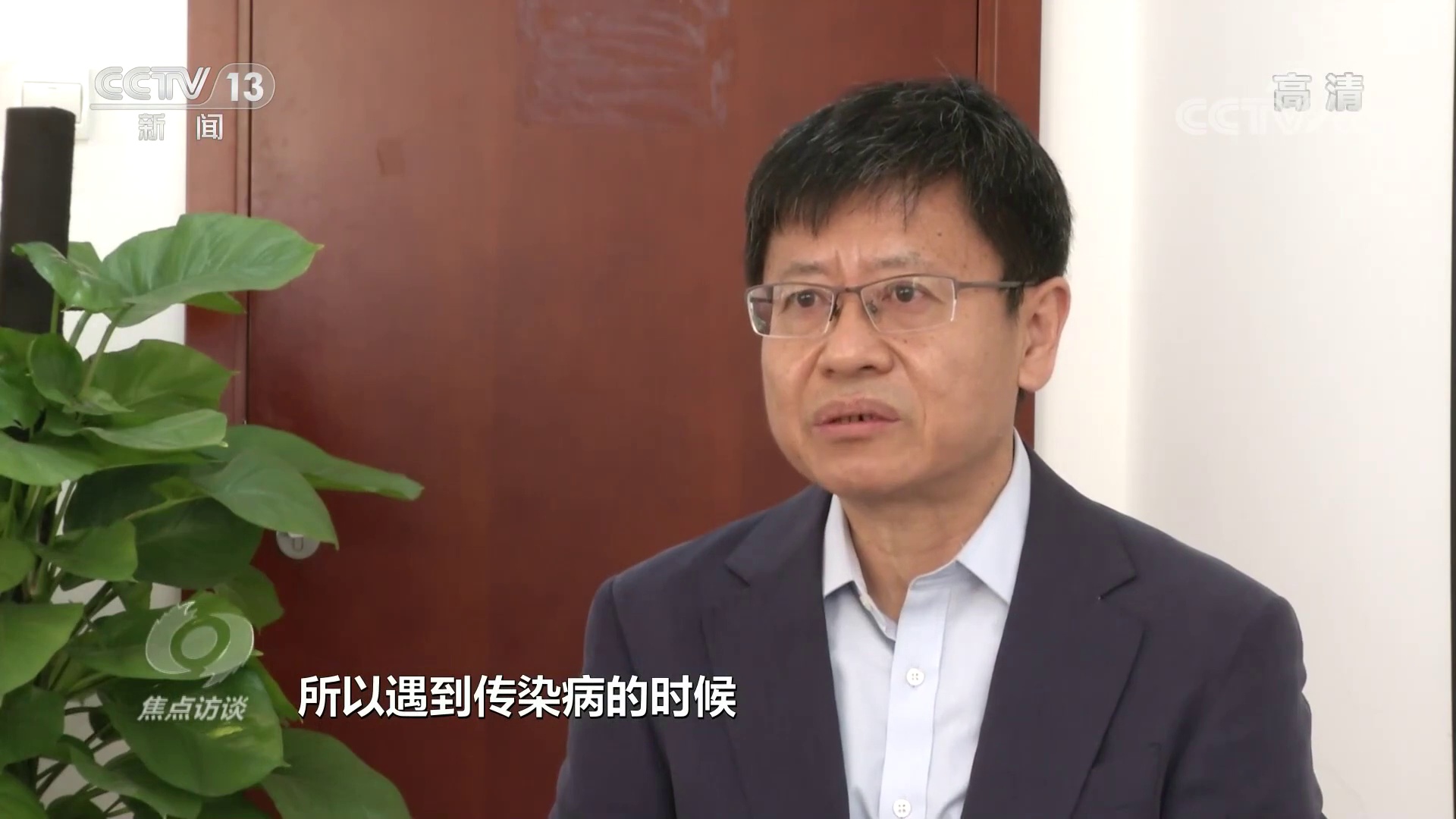}%
        \caption{COVID-19 (Chinese).}
        \label{subfig:cn_covid_77}%
    \end{subfigure}
    \hfill
    \begin{subfigure}{0.1\textwidth}
        \centering
        \includegraphics[scale=0.07]{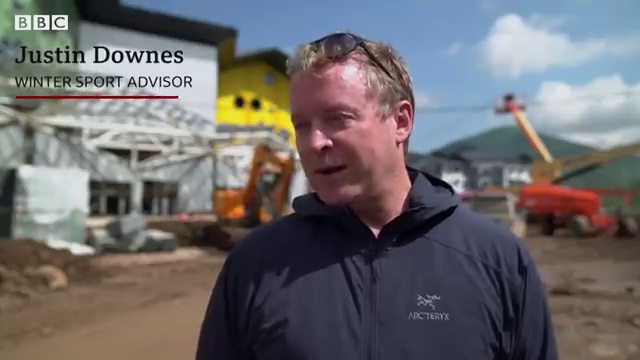}%
        \caption{Olympics (English).}
        \label{subfig:en_olympic_8}%
    \end{subfigure}
    \hfill
    \begin{subfigure}{0.1\textwidth}
        \centering
        \includegraphics[scale=0.07]{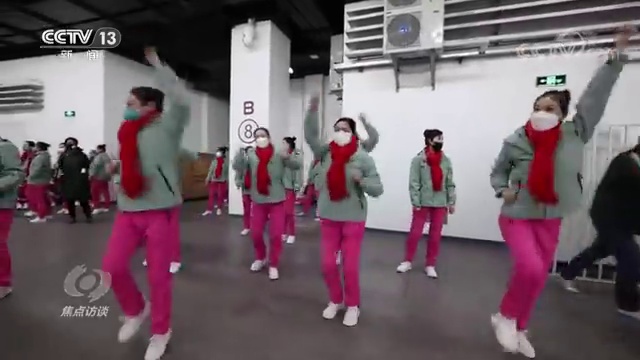}%
        \caption{Olympics (Chinese).}
        \label{subfig:cn_olympic_33}%
    \end{subfigure}

    \caption{Sample frames extracted.}
    \label{fig:sample}
    \Description{4 representative frames, each sampled from one video.}
\end{figure}

% original figure is 2 x 2
\begin{comment}
\begin{figure}[ht]
    \centering
    \begin{subfigure}{0.23\textwidth}
        \centering
        \includegraphics[scale=0.06]{docs/en_covid_15.jpg}%
        \caption{U.S. official in interview.}
        \label{subfig:en_covid_15}%
    \end{subfigure}
    \hfill
    \begin{subfigure}{0.23\textwidth}
        \centering
        \includegraphics[scale=0.06]{docs/cn_covid_77.jpg}%
        \caption{Chinese official talking.}
        \label{subfig:cn_covid_77}%
    \end{subfigure}

    \begin{subfigure}{0.23\textwidth}
        \centering
        \includegraphics[scale=0.18]{docs/en_olympic_8.jpg}%
        \caption{Host in construction site.}
        \label{subfig:en_olympic_8}%
    \end{subfigure}
    \hfill
    \begin{subfigure}{0.23\textwidth}
        \centering
        \includegraphics[scale=0.18]{docs/cn_olympic_33.jpg}%
        \caption{Chinese actresses rehearsing.}
        \label{subfig:cn_olympic_33}%
    \end{subfigure}

    \caption{Sample frames (top: COVID-19; left: English.)}
    \label{fig:sample}
    \Description{4 representative frames, each sampled from one video.}
\end{figure}
\end{comment}

\subsection{English COVID-19 “New Variant” Video}

This video\footnote{Video link: \url{https://www.youtube.com/watch?v=doP5UacBlt0}} is published by CBS Mornings on YouTube.
The main content concerns the resurgence of the Omicron variant and its spread across the US.
Dr. Agus also discusses reactions and restriction policies in large cities like New York and Los Angeles. The video length is 3 minutes and 40 seconds in total.

\subsubsection{Images.}

This video generates 109 image frames.
See Figure~\ref{subfig:en_covid_15} for a manually selected but typical frame: a shot of the host interviewing a medical expert talking to the camera.

\subsubsection{Text.}

The text data is collected from YouTube auto-generated closed captions in English, enabled by the Python package, \textsc{youtube-transcript-api}.
Since the autogenerated caption does not include punctuation, we use the Basic English Tokenizer in the PyTorch \textsc{torchtext} package.
We then use a pre-trained GloVe (“Global Vector”) Embedding with a dimension of 300 to embed every word token.

The \textsc{transcript} API segments the whole script paragraph into 96 segments of text, which gives approximately 10 to 20 words in each segment.
In addition, it also provides an accurate timestamp of each segment's beginning and end.
The period between two neighboring segments are roughly two seconds, which agrees with the rate at which we sample the frames.
We heuristically but automatically align frames to text segments by selecting the frame closest to a segment's starting time.

Sample text segments are given below:

\begin{verbatim}
   “health officials here in new york city”
   “and in los angeles are sounding the...”
\end{verbatim}

\subsection{Chinese COVID-19 “Vaccine” Video}

This video\footnote{Video link: \url{https://www.youtube.com/watch?v=xcWeBCOMoiU}} is published by China Central Television (CCTV) on YouTube.
The main content concerns an effort to encourage elderly Chinese citizens to take vaccines for COVID-19, while also reporting on the pandemic and the progress of disease control.
The rather long video length is 15 minutes and 8 seconds in total.

\subsubsection{Images.}

This video generates 378 image frames in total.
See Figure~\ref{subfig:cn_covid_77} for a manually selected but typical frame: a shot of a Chinese official talking.

\subsubsection{Text.}

Since YouTube does not provide transcripts for Chinese videos, we resort to YueLu Voice Club, a Chinese Automatic Speech Recognition converter.

However, some of the source video involves Chinese dialects somewhat different from Mandarin, so we manually corrected its mistakes.
We also added a timestamp every 10 seconds, a period chosen in consideration of the information density of Chinese relative to English.

This process gives us 90 text segments, which need to be aligned to the 378 image frames.
Since the average sample rate is 4.2 frames per segment, we heuristically but uniformly sample 5 images out of a series of 21 image frames, under the assumption that for news videos, Chinese words (or characters) are also uniformly distributed.

We used the Jieba Chinese word tokenizer as well as the Chinese Word Vectors (CWV) embedding \citep{chn_word2vec}, which embeds Chinese words and characters into vectors of dimension 300.

However, Jieba does not necessarily remove punctuation during tokenization, so we explicitly filter out Chinese-style punctuation like “,”, “.”, “!”, and “?”.
Additionally, some Chinese words as defined by Jieba (approximately 6\%), that are not recognized by CWV, e.g., the Chinese abbreviation of the word “coronavirus”.

Such words become "<UNK>" tokens (a vector of zeros), or are further split into single characters (e.g., the word “coronavirus” is split into “new” and “crown”.)
Sample text segments (translated to English) are given below:

\begin{verbatim}
    “Should the elderly people take COVID vaccines?
    Yes, elderly people must take vaccines. It can 
    prevent severe illness and deaths...”
\end{verbatim}

\subsection{English Olympic “Construction” Video}

This video\footnote{Video link: \url{https://www.youtube.com/watch?v=84rPLijOLyg}} is published by British Broadcasting Corporation (BBC) News on YouTube.
The main content concerns the construction and architecture of venues (e.g., winter sports centers) in Beijing for the Winter Olympics.
The video is relatively shorter, with the length of 2 minutes and 52 seconds.
We apply the same pre-processing steps as for English COVID-19 “New Variant” video.

\subsubsection{Images.}

This video generates 60 image frames.
See Figure~\ref{subfig:en_olympic_8} for a manually selected but typical frame: a Canadian host talking at a construction site.

\subsubsection{Text.}

Sample text segments are given below:

\begin{verbatim}
   “In the mountains around the Chinese capital,”
   “thousands of workers are busy...”
\end{verbatim}

\subsection{Chinese Olympic “Ceremony” Video}

This video\footnote{Video link: \url{https://www.youtube.com/watch?v=plAfxGnmadw}} is also published by China Central Television (CCTV) on YouTube.
The main content concerns the opening ceremony of Winter Olympics 2022, including the arrangements of the director, the practice and rehearsal of actresses, etc.
The video length is 15 minutes and 9 seconds in total.

\subsubsection{Images.}

Likewise, the same alignment steps are applied as for Chinese COVID-19 “Vaccine” images.
This video generates 91 image frames after sampling.
See Figure~\ref{subfig:cn_olympic_33} for a manually selected but typical frame: a group of Chinese actresses practicing for their dance show.

\subsubsection{Text.}

In additional to Chinese texts, this video also involves the speech in English by Yiannis Exarchos, the CEO of Olympic Broadcasting Services (OBS).
So the pre-processing step of the text combines both Chinese CWV embedding and English GloVe embedding.
Such text is embedded into numerical vectors before being passed to the CRVAE model, so the model is inherently multilingual.

Sample text segments (translated to English) are given below:

\begin{verbatim}
    “What surprises will be brought by the Feast of Snow?”
    “People perform warm-up shows in the way they love...”
\end{verbatim}

\section{Results}

\subsection{Autoencoder Model Experiments}

\subsubsection{Losses.}

We test the performance of the three CVAE variant models on the image encoding task using English and Chinese COVID-19 videos.
The losses are listed in Table~\ref{tab:cvae_variants}, where the MSE image losses have been properly scaled to account for differences in the image resolution within CRVAE.
We find that a dense CVAE is superior to a pure CVAE.
We note, however, that there is an increased training time cost.
Pure CVAE is trained for 100 epochs; dense CVAE for 300 epochs; and CRVAE for 500 epochs.

\begin{table}[ht]
    \centering
    \begin{tabular}{llll}
        \multicolumn{2}{c}{}    &\multicolumn{2}{c}{COVID-19 Data}\\
        \cmidrule(r){3-4}
        Model       &Loss Type  &EN                 &CN\\
        \midrule
        Pure CVAE   &Image      &0.3645             &0.1461\\
        Dense CVAE  &Image      &0.1653             &0.0955\\
        CRVAE       &Image      &\textbf{0.0408}    &\textbf{0.0323}\\
    \end{tabular}
    \caption{CVAE and CRVAE Model Performance.}
    \label{tab:cvae_variants}
\end{table}

Table \ref{tab:crvae_subvariants} is a comparison of performance across subvariants of our CRVAE model.
We find that, in COVID-19 datasets, the CRVAE model LSTM subvariant augmented with the Teacher Forcing algorithm is superior to the best RNN subvariant, and to the LSTM subvariant without Teacher Forcing.
Thus, we only apply this subvariant to the Winter Olympic datasets, and similar performances of image losses and text losses are observed on the two dataset pairs.

% Figure~\ref{fig:loss_crvae} plots the loss curves.
We note that text losses are significantly less than image losses.
This is due to their difference in dimension: texts have dimension $SeqLength \times EmbedSize \approx 10 \times 300 = 3,000$, while images have dimension $H \times W \times C = 200 \times 120 \times 3 = 72,000$.

\begin{table}[ht]
    \centering
    \begin{tabular}{llllll}
        \multicolumn{2}{c}{}    &\multicolumn{2}{c}{COVID-19 Data}      &\multicolumn{2}{c}{Olympic Data}\\
        \cmidrule(r){3-4}       \cmidrule(r){5-6}
        Subvariant  &Loss Type  &EN                 &CN                 &EN     &CN\\
        \midrule
        RNN + TF    &Total      &0.1064             &0.1002             &––     &––\\
        LSTM        &Total      &0.0520             &0.0539             &––     &––\\
        LSTM + TF   &Total      &\textbf{0.0441}    &\textbf{0.0340}    &0.0159 &0.0338\\
        LSTM + TF   &Image      &0.0408             &0.0323             &0.0129 &0.0311\\
        LSTM + TF   &Text       &0.0011             &0.0009             &0.0010 &0.0009\\
    \end{tabular}
    \caption{CRVAE Subvariants Model Performance.}
    \label{tab:crvae_subvariants}
\end{table}

\begin{comment}
\begin{figure}
    \centering
    \begin{subfigure}{0.23\textwidth}
        \centering
        \includegraphics[scale=0.23]{docs/loss_en_covid.png}
        \caption{English COVID-19 data.}
        \label{subfig:loss_en_covid}
    \end{subfigure}%
    \hfill
    \begin{subfigure}{0.23\textwidth}
        \centering
        \includegraphics[scale=0.23]{docs/loss_cn_covid.png}
        \caption{Chinese COVID-19 data.}
        \label{subfig:loss_cn_covid}   
    \end{subfigure}

    \begin{subfigure}{0.23\textwidth}
        \centering
        \includegraphics[scale=0.23]{docs/loss_en_olympic.png}%
        \caption{English Olympics data.}
        \label{subfig:loss_en_olympic}%
    \end{subfigure}
    \hfill
    \begin{subfigure}{0.23\textwidth}
        \centering
        \includegraphics[scale=0.23]{docs/loss_cn_olympic.png}%
        \caption{Chinese Olympics data.}
        \label{subfig:loss_cn_olympic}%
    \end{subfigure}
    \caption{CRVAE MSE training losses.}
    \label{fig:loss_crvae}
    \Description{4 loss curves, each reflects the training process of one dataset. All of them converge in less than 300 epochs, and stop at epoch 500.}    
\end{figure}
\end{comment}

\subsubsection{Image Reconstructions.}

\begin{figure*}[ht]
    \centering
        \begin{subfigure}{0.48\textwidth}
            \centering           
            \includegraphics[width=\linewidth]{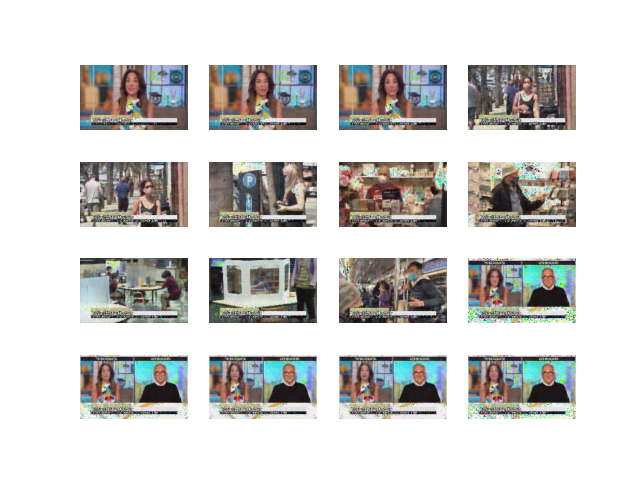}
            \caption{English data}
            \label{subfig:en_recon}
        \end{subfigure}
    \hfill
    \centering
    \begin{subfigure}{0.48\textwidth}
        \centering
        \includegraphics[width=\linewidth]{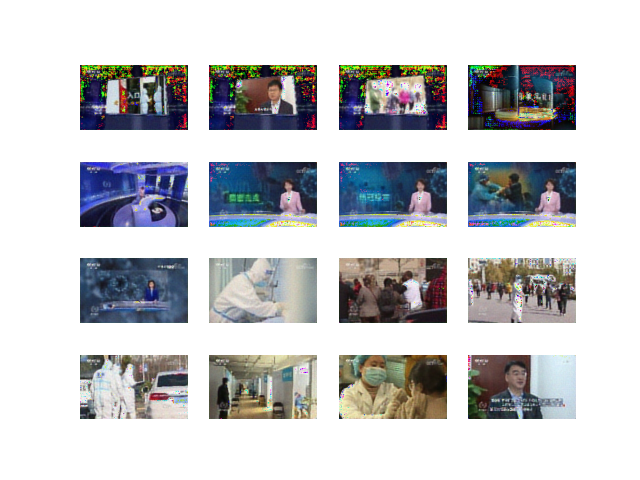}
        \caption{Chinese data.}
        \label{subfig:cn_recon}
    \end{subfigure}
    \caption{Sample reconstructed images, COVID-19 data.}
    \label{fig:sample_reconstructed}
    \Description{Each reconstructed image set contains $4 \times 4 = 16$ images extracted from the first batch of epoch 500. In general, the images are almost the same as the inputs, but some has bright spots (e.g., red, green, etc.) shown on white backgrounds.}
\end{figure*}

A sample of the reconstructed images of our model on the COVID-19 video pair are shown in Figure~\ref{fig:sample_reconstructed}; the Olympics video pair results are similar.
These images are generally clearer than those reconstructed by CVAE, which does not learn text simultaneously.

We noted a characteristic artifact in all CVAE-derived models.
If the original image has a light background, then the reconstructed image will likely have numerous small bright patches of saturated colors in those areas.
It appears that this is due to those pixels having a zero value in one or more of their RGB channels, although the root cause is still speculative.

\subsubsection{Text Reconstructions.}

\begin{figure*}[ht]
    \centering
    \includegraphics[width=\linewidth]{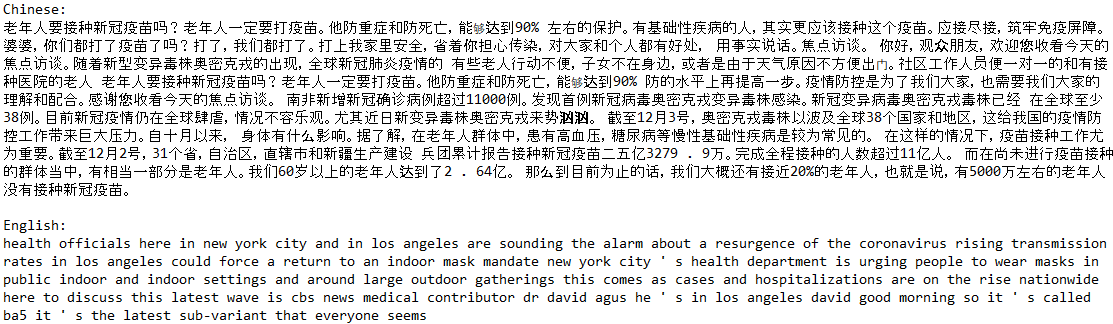}
    \caption{Sample reconstructed text for Chinese and English, COVID-19 data.}
    \label{fig:decode_text}
    \Description{The reconstructed English and Chinese text on epoch 500. The English text reads: “health officials here in New York City and in Los Angeles are sounding the alarm...”}
\end{figure*}

A sample of the reconstructed texts of our model, for both English and Chinese (COVID-19 pair), are shown in Figure~\ref{fig:decode_text}, after “verbalizing” the embeddings and cleaning up any padded tokens (as detailed in Section~\ref{subsec:decoder}).
We note that unknown tokens and filler words are often verbalized as “well” in English, because the method we use to decode embeddings into discrete words is by finding their nearest neighbors, and the embedding of “well” is close to the zero vector.
The results appear mostly accurate and properly formed.

\subsection{Clustering}

We use $K$-means to find vector clusters.
We manually examined the inter- and cross-cluster distances, as well as cluster size distribution.
Typical $K$'s are set to 3, 4, or 5, as shown in the Figures of Section~\ref{subsec:cross}.

\begin{figure}[ht]
    \centering
    \begin{subfigure}{0.23\textwidth}
        \centering
        \includegraphics[scale=0.23]{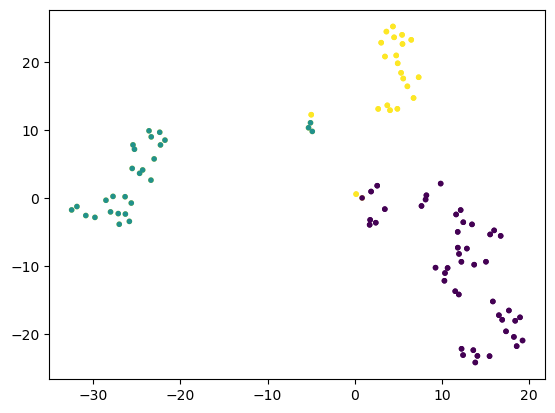}
    \end{subfigure}
    \begin{subfigure}{0.23\textwidth}
        \centering
        \includegraphics[scale=0.23]{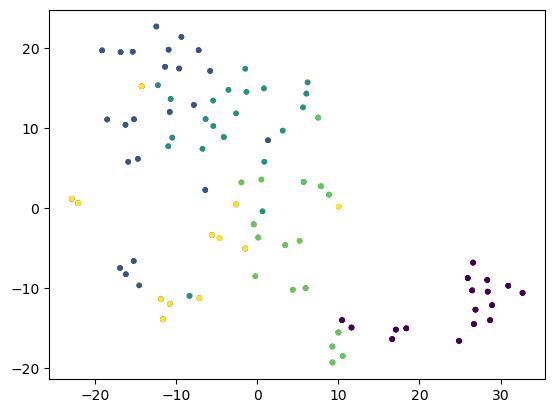}
    \end{subfigure}
    \caption{COVID-19 data t-SNE clusters (left: English).}
    \label{fig:tsne}
    \Description{Two t-SNE plots where points are shown in different colors that represent clusters. The left plot has 3 colors, while the right one has 5 colors.}
\end{figure}

The resulting clusters appear quite reasonable when projected into the 2D plane, using t-SNE~\cite{t-SNE}.  
We set the perplexity parameter to 8 for the COVID-19 datasets; see Figure~\ref{fig:tsne} (left) and (right) for the 3 English clusters, and the 5 Chinese, respectively.
Results for the Olympics data were similar: 4 English and 4 Chinese clusters.

\subsection{Cluster Interpretation}

To help understand the meanings of each cluster (COVID-19 data), we take advantage of three large models.

\subsubsection{BLIP-Generated Image Descriptions.}

Overall, the results of image descriptions generated by BLIP are satisfying. 
It can successfully detect persons, objects, backgrounds, etc., in a frame, and make inferences to some extent.
For example, BLIP detected the “laboratory” setting in the Chinese data (“A person in a lab holding a bottle”), and inferred the occasion of a TV show in English data (“A woman in a colorful dress is on the set of the Today show”). 

However, BLIP shows considerable disability in Optical Character Recognition tasks.
For example, it misreads the Chinese words “Priority pass for people aged 60+” into “No to the government”, which is nowhere in the image.
In English data, it misinterprets the text of a statistical table as a repetitive sequence of the words “corona” and “covids”.

\subsubsection{LLaMA-Generated Tags.}

We experimented with the tag generation process, 10 tags for each cluster, using Llama 2, among which 5 representative tags for each clusters are shown in Tables~\ref{tab:us_tags}~and~\ref{tab:cn_tags}.
We have manually traced back the originating medium of the output tags.  
Those tags marked by “$*$” and “$\dagger$” reflect information primarily in image frames and text captions, respectively.
(We did not label heavily shared common tags, such as “corona”.)

In either video, we find that most tags are meaningful, that is, they reflect actual information from images or texts.
Yet there are still about 20\% that contain generic, mistaken, or made-up information (hallucination).
Despite these inefficiencies, we found that the cluster interpretation pipeline significantly reduced human effort in exploring the meaning of a cluster.

We verified that the results with 3 clusters are most sensible for the English COVID-19 video, as shown in Table \ref{tab:us_tags}, while those with 5 clusters are most sensible for the Chinese COVID-19 video, as shown in Table \ref{tab:cn_tags}.
We similarly verified the results for the Olympics videos.
Overall, in keeping with the clustering results, the LLaMA model performed better on the English video than the Chinese video, most likely in part because the training corpus of Llama 2 is mainly in English.

\begin{table*}[ht]
    \centering
    \begin{tabular}{lll}
        Impacts and strategies           &Talking head        &Scientific analytics\\
        \midrule
        New COVID-19 variant identified$^\dagger$   &Medical Contributor$^\dagger$  &covid-19 on the rise$^\dagger$\\
        Indoor mask mandate possible$^\dagger$      &Repeated Illnesses$^\dagger$   &green and white circle with words$^*$\\
        Living with the virus$^\dagger$             &Women$^*$                      &group of people sitting at a table$^*$\\
        Health officials sound the alarm$^\dagger$  &Dress$^*$                      &percentage of coronavirus in the US$^*$\\
        Ba.2.75: the latest COVID-19 wave           &Ba5 Variant                    &corona corona seen in graphic\\
    \end{tabular}
    \caption{Sample tags of English COVID video (3 Clusters).}
    \label{tab:us_tags}
    \centering
    \begin{tabular}{lllll}
        Talking head            &Global response            &Treatment          &Professional comments  &Elderly vaccine\\
        \midrule
        Health concerns$^\dagger$   &Global Response$^\dagger$      &Risk$^\dagger$         &Virus$^\dagger$    &Older Adults$^\dagger$\\
        Global pandemic$^\dagger$   &Infection Control$^\dagger$    &Immunity$^\dagger$     &Immune System$^\dagger$    &Government$^\dagger$\\
        Hospital scene$^*$          &Patient$^*$                    &Vaccine$^\dagger$      &Elderly$^\dagger$          &Social Responsibility$^\dagger$\\
        TV host$^*$                 &Medical Staff$^*$              &Protection$^\dagger$   &Medical Professional$^*$   &Medical Examinations$^*$\\
        Dental care                 &Quarantine                     &COVID-19               &Pandemic                   &Influenza\\
    \end{tabular}
    \caption{Sample tags of Chinese COVID video (5 clusters).}
    \label{tab:cn_tags}
    \begin{tabular}{c}
         Tags labeled “$*$” and “$\dagger$” are from image frames and text captions, respectively.
    \end{tabular}
\end{table*}

% EXPERIMENT TO SEE IF FUTURE WORK TABLE FITS BETTER
\begin{table*}[hbt]
    \centering
    \begin{tabular}{lll}
        Topic                           &English video focus                        &Chinese video focus\\
        \midrule
        China visa-free policy          &Convenience; EU government responses       &Tourists bring business opportunities to China\\
        Beryl Hurricane                 & Impact to the Hurricane season ahead      &Other extreme weathers in the U.S.\\
        Copa America 2024               &Delayed kick-off due to chaos              &Argentina football fans; Messi's performance\\
        China self-driving robotaxi     &Comparison to Tesla’s Robotaxi             &Tech issues; impact to taxi drivers' employment\\
        China contaminated cooking oil  &Chinese media’s reactions to the scandal   &Who should take the responsibility
    \end{tabular}
    \caption{Expected differences in cluster outcomes between cultures of five additional topics, based on human inspection.}
    \label{tab:five_topics}
\end{table*}

\subsubsection{PhraseBERT-Embedded Vectors}

Having summarized the content in English tags, we use \textsc{phrase-bert} \cite{phrasebertwang2021} to measure tag similarities in the BERT 768-dimensional space.
(We find that GloVe embeddings are less complex, and LLaMA-based embeddings are designed primarily for text generation tasks.)

\subsection{Cross-cultural Comparisons}
\label{subsec:cross}

\begin{figure}[ht]
    \centering
    \includegraphics[scale=0.4]{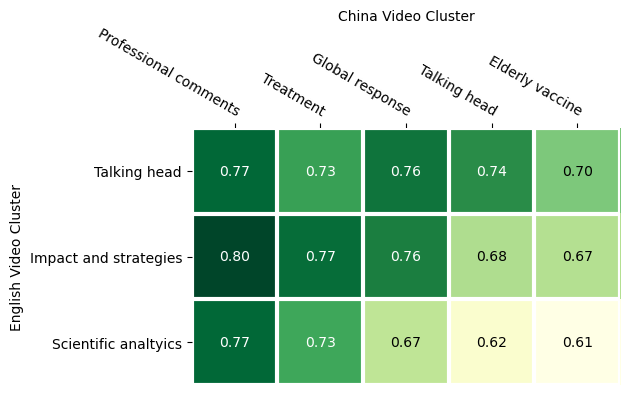}
    \caption{Similarities of COVID-19 cluster pairings.}
    \label{fig:grouped_cos_sim_covid}
    \Description{A heatmap of $3 \times 5$, where the top-left corner shows clusters with high similarity, while bottom-right corner shows those with low similarity. The similarities range from 0.61 to 0.80.}
\end{figure}

\begin{figure}[ht]
    \centering
    \includegraphics[scale=0.4]{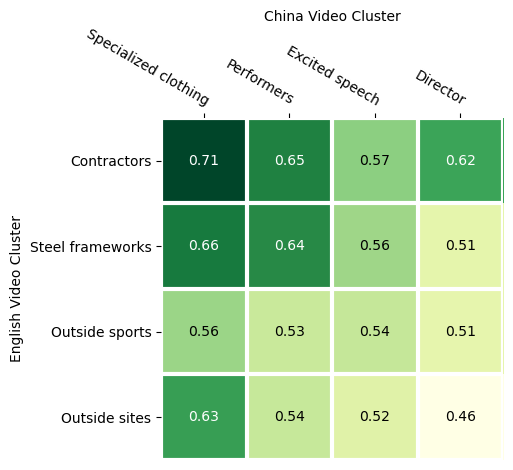}
    \caption{Similarities of Olympics cluster pairings.}
    \label{fig:grouped_cos_sim_olympic}
    \Description{A heatmap of $4 \times 4$, where the top-left corner shows clusters with high similarity, while bottom-right corner shows those with low similarity. The similarities range from 0.46 to 0.71.}
\end{figure}

We visualized the relation between clusters across cultures in Figure~\ref{fig:grouped_cos_sim_covid}~and~\ref{fig:grouped_cos_sim_olympic}.
We encoded the tags using \textsc{phrase-bert} and computed a $10 \times 10$ cosine similarity matrix for each <English, Chinese> cluster pair.
The cluster pair similarity is measured by the average of top 10 cosine similarities within the cluster pair.
Further, the clusters are sorted in descending similarity order.
For convenience of further discussion, we have manually named each cluster by a summarizing phrase.

For the COVID data in Figure~\ref{fig:grouped_cos_sim_covid}, we ignore the requisite "Talking head" clusters.
We note that the English cluster “Impacts and strategies” is an overall summary of current pandemic status and vaccine research progress, which shares many similarities with the Chinese clusters “Global response”, “Treatment”, and “Professional comments”.
Those three Chinese clusters, in turn, have high similarity with almost all English clusters; in particular, the Chinese cluster “Professional comments” is especially related to English cluster “Scientific analytics”, with statistics and data visualizations.

Yet, Chinese cluster “Elderly vaccine” is quite distant from all the English clusters.
This identifies the Chinese-specific characteristics of the video: the emphasis on older people and on the social responsibility of taking vaccines.

In the Olympic data in Figure~\ref{fig:grouped_cos_sim_olympic}, we observe high similarities between the English “Contractors” and “Steel frameworks” clusters and the Chinese “Specialized clothing” and “Performers” clusters.
These two pairs of clusters depict China's preparation for the Winter Olympics from different aspects.
The English video emphasizes the “hardware” side, i.e., the construction of venues, with two typical scenes being contract workers on the construction site, and the buildings that are in progress.
The Chinese video emphasizes the “software” side, i.e., actors and actresses practicing for the spectacle of the opening ceremony, with people in specialized costumes being interviewed.

In contrast, the other two English and Chinese clusters are rather distant, especially the English “Outside sites” cluster and Chinese “Director” cluster, which occurs indoors.

\section{Discussion and Future Work}

% NOTE: the following command "balances" the number of lines in the two columns of the last page.  But it still might trigger an error, since the ACM style sheet is broken.  See https://tex.stackexchange.com/questions/504955/package-balance-warning-in-acmart
% \balance

% ORIGINAL LOCATION OF FUTURE WORK TABLE
\begin{comment}
\begin{table*}[hbt]
    \centering
    \begin{tabular}{lll}
        Topic   &English video  &Chinese video\\
        \midrule
        China visa-free policy       &Convenience; EU government responses             &Tourists bring business opportunities to China\\
        Beryl Hurricane       & Impact to the Hurricane season ahead              &Other extreme weather in US\\
        Copa America 2024&Delayed kick-off due to chaos&Argentina football fans; Messi's performance\\
        China self-driving robotaxi&Comparison between Baidu’s and Tesla’s&Tech issues; taxi drivers' potential unemployment\\
        China contaminated cooking oil&Chinese media’s reactions to the scandal&Who should take the responsibility
    \end{tabular}
    \caption{Caption}
    \label{tab:five_topics}
\end{table*}
\end{comment}

We have fully implemented our system, with two videos each (Chinese and English) on two very different topics, COVID-19 and Winter Olympics.
As discussed in Section~\ref{sec:dataset},
we are applying our pipeline to five additional topics; see Table~\ref{tab:five_topics} for details.
Additionally, we 
will be extending some topics to include additional videos.

In this work, we constructed a full pipeline for integrating and processing video data through different modalities: images, audio, and text.
Our Convolutional-Recurrent Variational Autoencoder architecture, CRVAE, combined the strengths of CVAE and LSTM models, while maintaining the encoder-decoder structure of both, and demonstrating correctness and superiority.

We adapted pre-trained Large Vision-Language Models and Large Language Models (Llama 2, BLIP, and PhraseBERT) to add explainability to the clustering process, with neither ground-truth nor supervision, through the generation of high-quality, human-interpretable tags for the resulting video clusters.

Additionally, we showed how these tools not only compactly captured the thematic content of the video pairs, but also provided an intuitive visualization of the primary differences between the pairings.
Although we illustrated our system with two specific news occurrences as covered by two specific viewpoints, we believe the methods are easily extensible for summarizing and contrasting most other events and cultural styles.

More technically, our work suggests future research areas, such as replacing our LSTM model by a Transformer model, or augmenting other models with a similar two-network architecture (e.g., CGAN, Convolutional Generative Adversarial Network) with LSTMs or Transformers.

% \subsection{Acknowledgments} We would like to express our gratitude toward fellow researcher Omer Onder, Alan Luo, Zheng Hui, and Hui's former teammate Zihang Xu at Columbia University. Luo has contributed to the web-scraping of image data, and Hui shared with us his experience of Chinese word tokenizer and embedding. Omer Onder provided an example of a Pure CVAE.

% \subsection{Downloads} The project's source code is available on GitHub \url{https://github.com/Anemonee1212/cvae_video_cluster} and \url{https://github.com/Anemonee1212/crvae_video_cluster}.

% ==================================================
\clearpage
\balance
\bibliographystyle{ACM-Reference-Format}
\bibliography{reference}

% \clearpage
\section{Appendix}

\subsection{BLIP Caption Prompt} \label{sec:blip_prompt}

\begin{samepage}
We experimented with both unconditional and conditional (i.e., prompted) BLIP captioning.
The prompt for conditional BLIP captioning is given below:
\begin{verbatim}
   A news photo of ... {BLIP will generate description}
\end{verbatim}
\end{samepage}

\subsection{LLaMA Prompt} \label{sec:llama_prompt}

\begin{verbatim}
   <s> [INST] <<SYS>>
   Please generate 10 short tags for a series of
   frames sampled from a YouTube news video based on
   the images and captions provided.  Please avoid 
   generic words that describe the whole video, but
   emphasize the unique characteristics of these 
   frames.  You may need to implicitly infer the
   meanings of the objects in the image description
   according to the video context.
   <</SYS>>
   Text caption: {inputs in English or Chinese...}
   Image description: {BLIP inputs in English...}
   [/INST] </s>
\end{verbatim}

The purpose of the second sentence is to avoid common keywords that appear in all clusters (e.g., “coronavirus” or “Olympics”).
The purpose of the last sentence is to emphasize the contextual understanding between image frames and text captions.
As an example, a BLIP description can be “a man in a mask and protection suit”, and we want LLaMA to infer that this man is likely a medical staff officer performing examinations for COVID-19.

The sampling process during LLaMA text generation is controlled by several hyperparameters, including temperature, $t$.
We experimented with different values of $t \in [0, 1]$.
With small $t$, LLaMA tends to respond in a conservative way, predicting text that most likely to follow, while with large $t$, it tends to be creative.
In practice, we set a large value of $t = 0.9$ for a diversified generation.

\end{document}